\pdfoutput=1
\documentclass[sigconf]{acmart}
\usepackage{pifont}
\usepackage[linesnumbered,ruled,vlined]{algorithm2e}
\usepackage{enumitem}
\newcommand{\cmark}{\ding{51}}%
\newcommand{\xmark}{\ding{55}}%
\usepackage{graphicx}
\usepackage{amsmath}

\setcopyright{acmcopyright}
\copyrightyear{2021}
\acmYear{2021}

\acmConference[Gold Coast '21]{Gold Coast '21: ACM Multimedia Asia}{December 01--03, 2021}{Gold Coast, Australia}
\acmBooktitle{Gold Coast '21: ACM Multimedia Asia,
  December 01--03, 2021, Gold Coast, Australia}

\begin{document}

\title{NoisyActions2M: A Multimedia Dataset for Video Understanding from Noisy Labels}


\author{Mohit Sharma}
\email{mohits@iiitd.ac.in}
\affiliation{
  \institution{Indraprastha Institute of Information Technology Delhi}
  \country{India}
}

\author{Raj Patra}
\authornote{Both authors contributed equally to this research.}
\email{rpatra1267@gmail.com}
\affiliation{
  \institution{National Institute of Technology Rourkela}
  \state{Odisha}
  \country{India}
}
\author{Harshal Desai}
\authornotemark[1]
\email{harshaldesai01@gmail.com}
\affiliation{
  \institution{National Institute of Technology Jamshedpur}
  \state{Jharkhand}
  \country{India}
}

\author{Shruti Vyas}
\email{shruti@crcv.ucf.edu}
 \affiliation{
  \institution{University of Central Florida}
  \city{Orlando}
  \country{Florida}
}

\author{Yogesh Rawat}
\email{yogesh@crcv.ucf.edu}
 \affiliation{
  \institution{University of Central Florida}
  \city{Orlando}
  \country{Florida}
}

\author{Rajiv Ratn Shah}
\email{rajivratn@iiitd.ac.in}
\affiliation{
  \institution{Indraprastha Institute of Information Technology Delhi}
  \country{India}
}







\begin{abstract}
    Deep learning has shown remarkable progress in a wide range of problems. However, efficient training of such models requires large-scale datasets, and getting annotations for such datasets can be challenging and costly. In this work, we explore the use of user-generated freely available labels from web videos for video understanding. We create a benchmark dataset consisting of around 2 million videos with associated user-generated annotations and other meta information. We utilize the collected dataset for action classification and demonstrate its usefulness with existing small-scale annotated datasets, UCF101 and HMDB51. We study different loss functions and two pretraining strategies, simple and self-supervised learning. We also show how a network pretrained on the proposed dataset can help against video corruption and label noise in downstream datasets. We present this as a benchmark dataset in noisy learning for video understanding. The dataset, code, and trained models will be publicly available for future research.
\end{abstract}



\maketitle

\section{Introduction}
The ImageNet dataset \cite{deng2009imagenet} has been one of the catalysts behind the exponential growth in Deep Learning \cite{lecun2015deep} and large scale machine learning research, along with transfer learning adoption to adapt large trained networks on problems with little data. This has led to many large-scale datasets targeting various tasks such as classification, detection, segmentation, etc., and a rising interest in training bigger networks to capture more variations and transfer well. ImageNet \cite{deng2009imagenet}, and Youtube8M \cite{abu2016youtube} are enormous datasets in terms of size and annotations. Still, it is not always possible to construct such massive annotated datasets due to logistical and time constraints.

Collecting data from the web is getting much popularity due to its availability on several social media platforms (e.g., Webvision \cite{li2017webvision}, and Clothing1M \cite{xiao2015learning}). Along with these datasets, many other works \cite{divvala2014learning} \cite{chen2015webly} have shown how learning from web data dramatically increases performance in related domains, despite labels which are mostly inferred from surrounding meta data and not manually verified. Moreover, the meta data itself acts as a rich source of information about the data point for tasks like image captioning, video understanding, etc.

As an active research area, there is a dire need to set a standard benchmark for efficient learning from noisy web data. With this objective in mind, we construct such a dataset with a primary focus on video modality. Our dataset consists of raw videos collected from Flickr, with surrounding meta data such as title, description, comments, etc. It has been collected using class labels from popular video classification benchmarks as search queries since these datasets have already established useful labels based on various criteria. We first look at various statistics of our dataset to set its importance and show some preliminary results for video action classification.

A major challenge that one encounters while learning from web collected data is its heavily imbalanced multi-label nature. Similar works \cite{ghadiyaram2019large} randomly select one label from the list of given multi labels. We compare various multi-label learning strategies in literature while pretraining on our dataset and also look at the setting of simple pretraining or combining it with self-supervised learning at various stages, hoping that this will set a benchmark for the research community. Finally, we also obtain some surprising results around how models pretrained on the proposed noisy dataset provide some robustness against label noise and video corruption with just simple fine-tuning and no modification to the training pipeline.

We first talk about related work in Section $2$. Next, in Section $3$, we discuss our dataset construction and statistics. We describe our methodology in Section $4$ and our experimental setup in Section $5$. We finally present our results and a discussion around them in Section $6$. We end with Section $7$, discussing how this work can be further improved and new research directions from our proposed dataset.

\section{Related Work}

\subsection{Video Datasets}


The UCF101 \cite{soomro2012ucf101} and the HMDB51 \cite{kuehne2011hmdb} datasets were one of the first datasets to spark interest in video understanding. However, they are relatively small for training large networks and required a lot of annotation effort. To this end, Heilbron et al. \cite{caba2015activitynet} proposed ActivityNet, which covered many common human activities and relatively longer videos. Along the same lines, Sigurdsson et al. \cite{sigurdsson2016hollywood} released the Charades dataset with a focus on everyday household activities. Kay et al. \cite{kay2017kinetics} introduced the Kinetics dataset with a primary focus only on covering a broad range of human activities and had a much larger number of videos than its contemporaries. Gu et al. \cite{gu2018ava} proposed the AVA dataset, which densely annotates $80$ `atomic' actions using movies. Goyal et al. \cite{goyal2017something} introduced the Something-Something dataset, where classes were defined as caption templates to enable solutions towards common sense understanding in videos. Finally, Zhao et al. \cite{zhao2019hacs} proposed the HACS dataset for action recognition and temporal localization, and models trained on this dataset showed excellent transfer learning performance. 

Karpathy et al. \cite{karpathy2014large} released a million scale weakly annotated dataset centered around sports. A similar scale dataset, proposed by Monfort et al. \cite{monfort2019moments} emphasized on event understanding. Diba et al. \cite{hvu2019} presented the HVU dataset, which is a multi-label dataset organized hierarchically in a semantic taxonomy, and constructed from Kinetics-600 \cite{kay2017kinetics}, Youtube-8M \cite{abu2016youtube} and the HACS dataset \cite{zhao2019hacs}. And lastly, Abu et al. \cite{abu2016youtube} proposed the Youtube-8M dataset, which is the biggest multi-label dataset for video understanding but only provides frame-level features. These datasets involved manual annotation of samples that differentiate them from the proposed dataset where the labels are inferred from the associated meta-data or corresponding search query, which does not require manual curation.


\begin{table*}[t]
\centering
\small
\caption{\small A comparison of various datasets used for video understanding.}
\label{tab:datasets}
\begin{tabular}{|c|c|c|c|c|c|c|c|c|}
  \hline
  \small Dataset & \# Classes & \# Videos & Noisy & Meta Data & Multi-Label & Source & Year
  \\
  \hline
  \small HMDB51 \cite{kuehne2011hmdb} & 51 & 7k & \xmark & \xmark & \xmark & Many & 2011 \\
  \small UCF101 \cite{soomro2012ucf101} & 101 & 13k & \xmark & \xmark & \xmark & Youtube & 2012 \\
  \small Sports-1M \cite{karpathy2014large} & 487 & 1M & \cmark & \xmark & \xmark & Youtube & 2014 \\
  \small ActivityNet \cite{caba2015activitynet} & 200 & 20k & \xmark & \xmark & \xmark & Youtube & 2015 \\
  \small Charades \cite{sigurdsson2016hollywood} & 157 & 10k & \xmark & \xmark & \xmark & Crowdsourced & 2016 \\
  \small Youtube-8M \cite{abu2016youtube} & 4800 & 8M(features) & \cmark* & \xmark & \cmark & Youtube & 2016 \\
  \small Kinetics \cite{kay2017kinetics} & 600 & 500k & \xmark & \xmark & \xmark & Youtube & 2017 \\
  \small Something-Something \cite{goyal2017something} & 174 & 108k & \xmark & \xmark & \xmark & Crowdsourced & 2017 \\
  \small AVA \cite{gu2018ava} & 80 & 576k & \xmark & \xmark & \xmark & Youtube & 2018 \\
  \small HACS \cite{zhao2019hacs} & 200 & 140k & \xmark & \xmark & \xmark & Youtube & 2019 \\
  \small HVU \cite{hvu2019} & 3457 & 572k & \xmark & \xmark & \cmark & Many & 2019 \\
  \small Moments in Time \cite{monfort2019moments} & 339 & 1M & \xmark & \xmark & \xmark & Many & 2019 \\
  \hline
  \small  \textbf{Ours: NoisyActions2M} & \textbf{7000} & \textbf{2M} & \textbf{\cmark} & \textbf{\cmark} & \textbf{\cmark} & \textbf{Flickr} & \textbf{2021} \\
  \hline
\end{tabular}\\
\footnotesize{* - Label Vocabulary constructed using Knowledge Graph API and human raters.}
\end{table*}

\begin{figure}
    \begin{center}
       \includegraphics[width=\linewidth, scale=0.5]{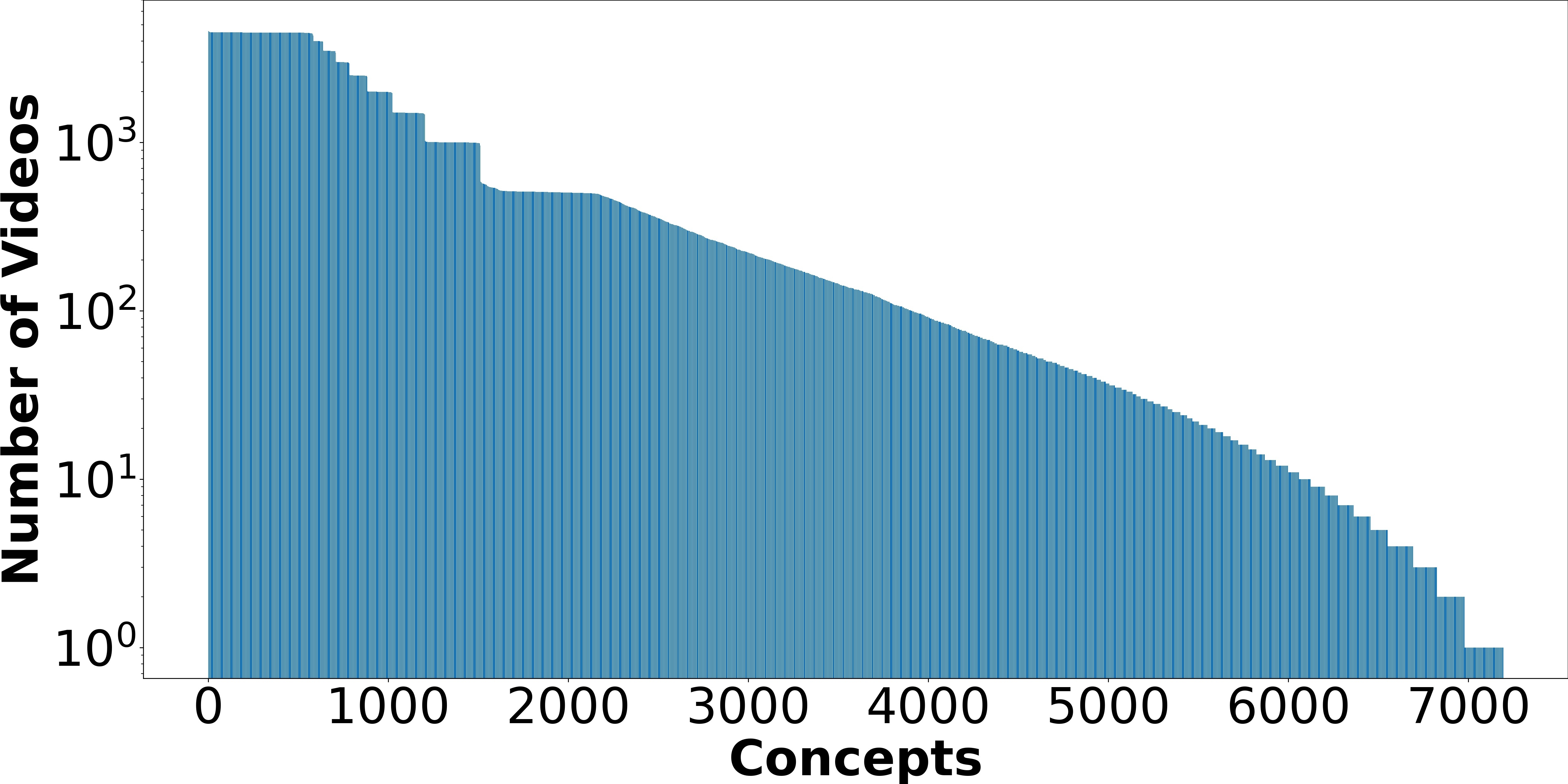}
    \end{center}
  \caption{Distribution of labels and videos (Log-Scale).}
  \label{fig:avgSamples}
\end{figure}

\subsection{Learning from Web Data}


Training data collected from the web is often used directly for some objective with weak labels. Weak labels are inferred from the surrounding text or meta data and are not verified, like Webvision \cite{li2017webvision}, and Clothing-1M \cite{xiao2015learning}. They also have a clean validation set for benchmarking methods used to learn from it. Also referred to as webly supervised learning, this area has been explored thoroughly in the literature. Divvala et al. in \cite{divvala2014learning} proposed a system to learn detectors for a given concept using different attributes from web data. Chen et al. in \cite{chen2015webly} presented a two-stage curriculum training, where they first learned from an easy set of images scraped from Google and then trained using images from Flickr.

Mahajan et al. \cite{mahajan2018exploring} trained an image classifier to predict hashtags on millions of social media images and observed huge gains in accuracies when the trained network is fine-tuned to various downstream datasets and tasks. Thomee et al. \cite{thomee2016yfcc100m} proposed the YFCC100M dataset with the intention of large-scale multimedia research and released $800k$ videos with general entities, along with $99M$ images and their meta data. A work very similar to ours is by Ghadiyaram et al. \cite{ghadiyaram2019large}, where they showed similar pretraining results for videos for improving downstream performance on action recognition along with interesting related experiments, but do not make their dataset public. Since there is a rising interest in research areas along these lines, our dataset will act as a benchmark to compare approaches to learn in webly supervised settings. Unlike other video datasets in this space, we also provide a rich set of meta data information corresponding to each video so that this dataset can be utilized for a variety of webly supervised training tasks beyond our current considered scope.


\section{Dataset Construction And Analysis}

\begin{figure}
    \begin{center}
       \includegraphics[width=\linewidth, scale=0.5]{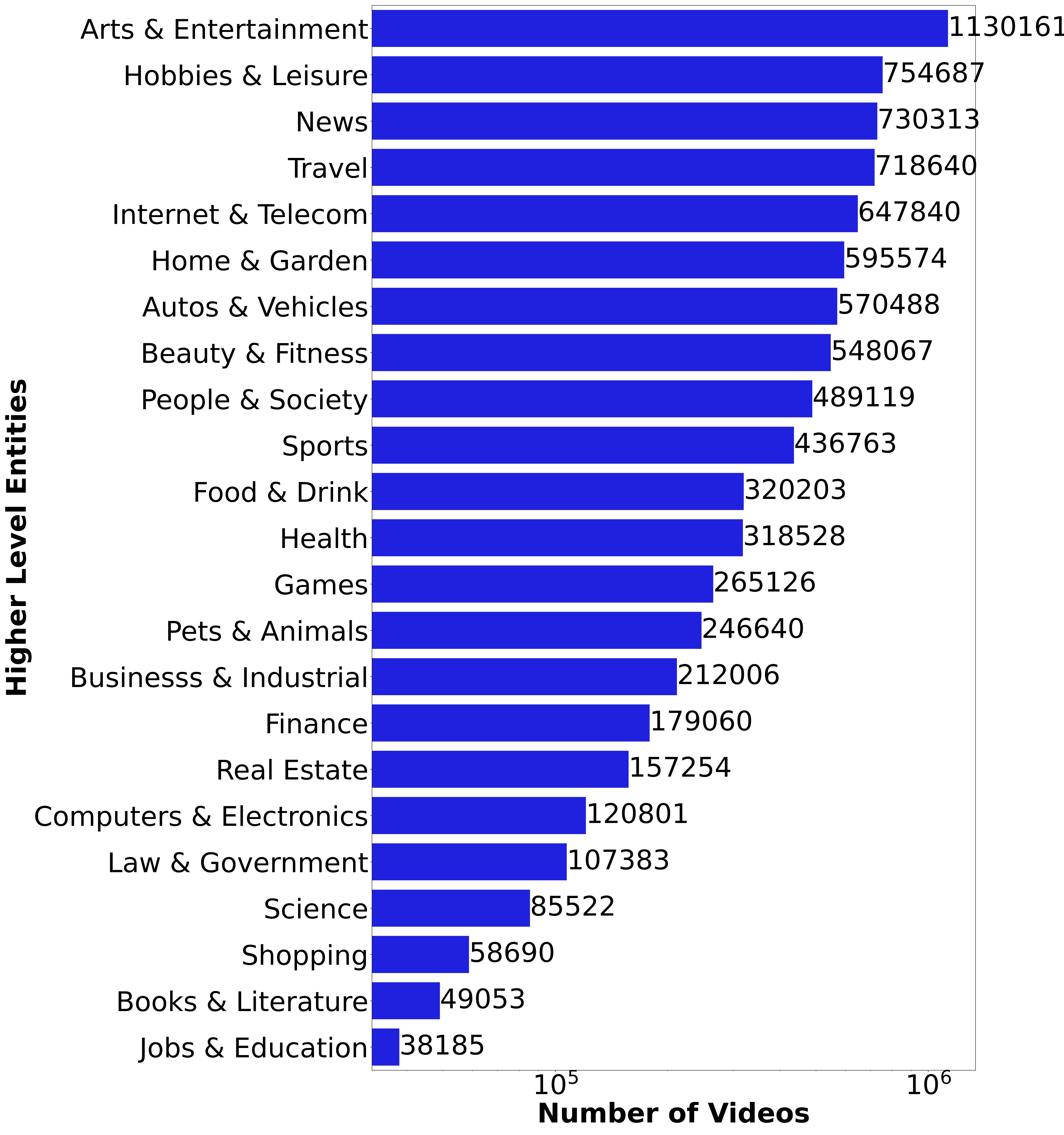}
    \end{center}
  \caption{Different types of Entities present in the dataset.}
  \label{fig:entityDistribution}
\end{figure}

\begin{figure*}
    \begin{center}
       \includegraphics[width=\linewidth, scale=0.5]{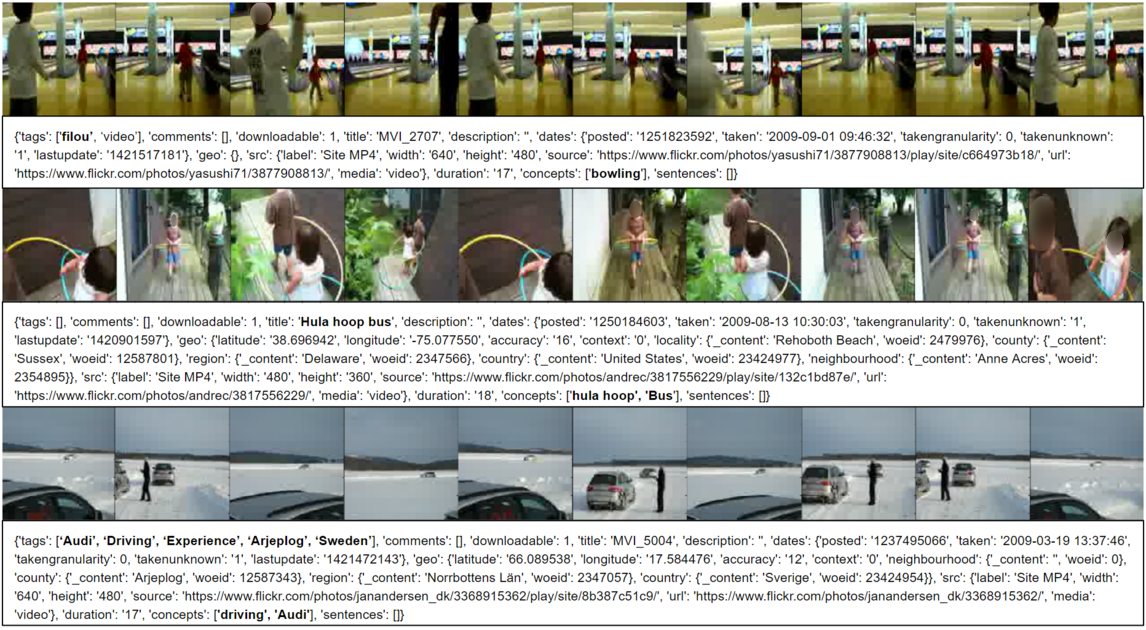}
    \end{center}
  \caption{Frames from some videos in NoisyActions2M, along with their meta data. Relevant meta information is highlighted in bold fonts.}
  \label{fig:frames}
\end{figure*}

\begin{figure}
    \begin{center}
       \includegraphics[width=\linewidth, scale=0.5]{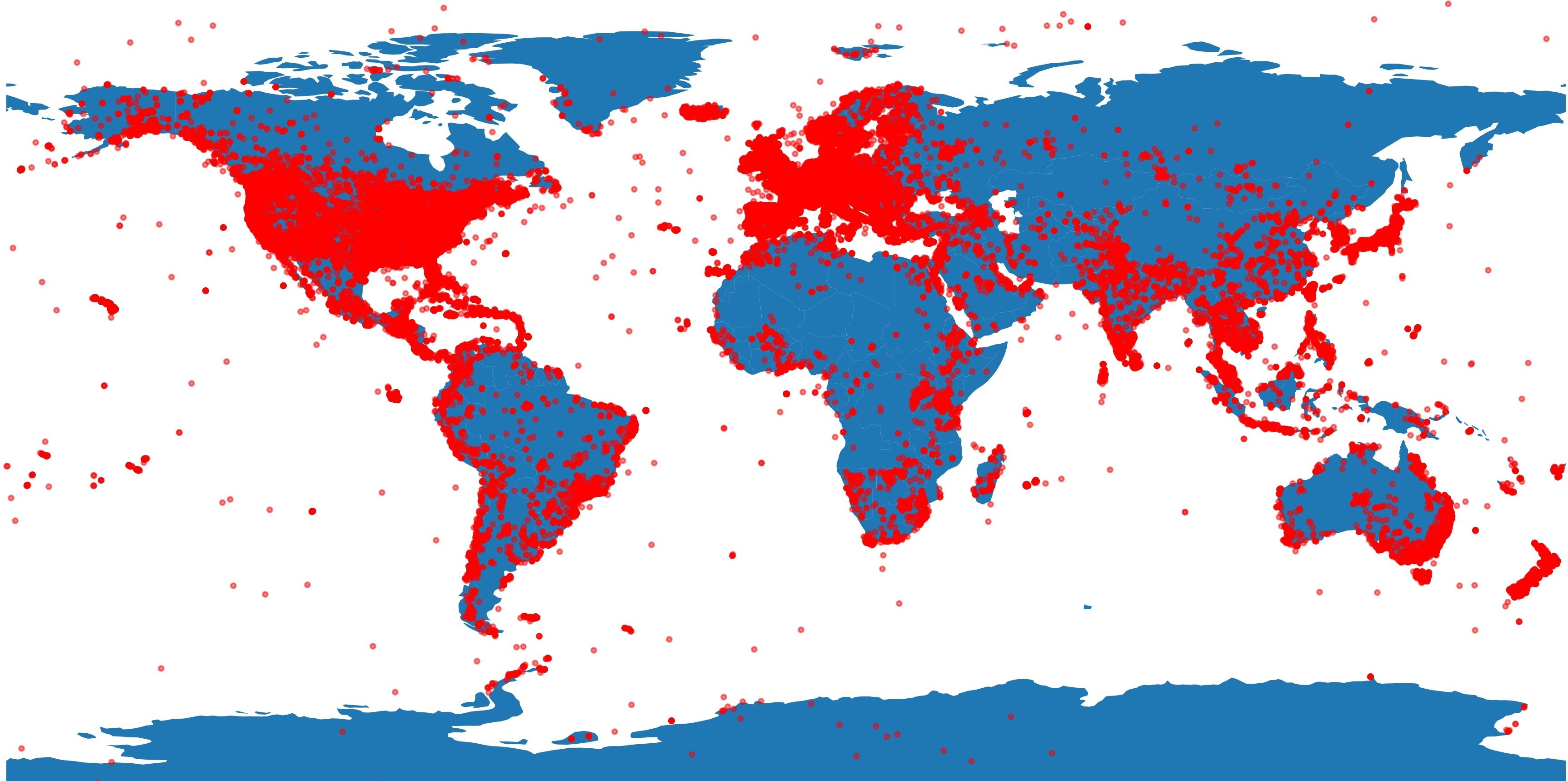}
    \end{center}
  \caption{Country Distribution of videos in the dataset}
  \label{fig:countryDistribution}
\end{figure}

We use the class labels from existing datasets which are listed in Table ~\ref{tab:datasets}) as search queries and collect the videos from Flickr. We lemmatize all the action class names, which results in some duplicates between sets of classes in different datasets, and only consider classes with more than $10$ videos. The total number of classes amounts to around $7000$, corresponding to roughly $1.95$ million videos. Around $0.05$ million videos are also collected using long sentence queries, which are lemmatized captions taken from various caption datasets, listed in Appendix A.1. The resulting class distribution in log10 scale is shown in Figure \ref{fig:avgSamples}.

For each video ID, we have the following information: tags (user assigned meta tags), comments, title, description, dates, geolocation, source (URL of the video with different sizes), duration, sentences (preprocessed long sentence queries), and concepts (the query action classes). Table \ref{tab:datasetStats} gives an overall picture of our datasets in terms of statistics (rounded off to nearest integer), demonstrating the various amounts of useful information this dataset will be able to provide. To look at the type of metadata present with the videos, we try to classify the title and concepts into some common types of entities. The details of the classification process are available in Appendix A.2. Figure \ref{fig:entityDistribution} shows the distribution of videos, which indicates the presence of diverse topics in the meta data with the videos. 

We also look at the regional diversity of our dataset, extracted using the geolocation attached with each video, shown in red in Figure \ref{fig:countryDistribution}. While many videos are from the North American and the European regions, other regions also have some representation. The Non-Popularity of Flickr in other regions might be one reason for not getting a large number of video samples from those regions.

\begin{figure*}
    \begin{center}
       \includegraphics[width=\linewidth]{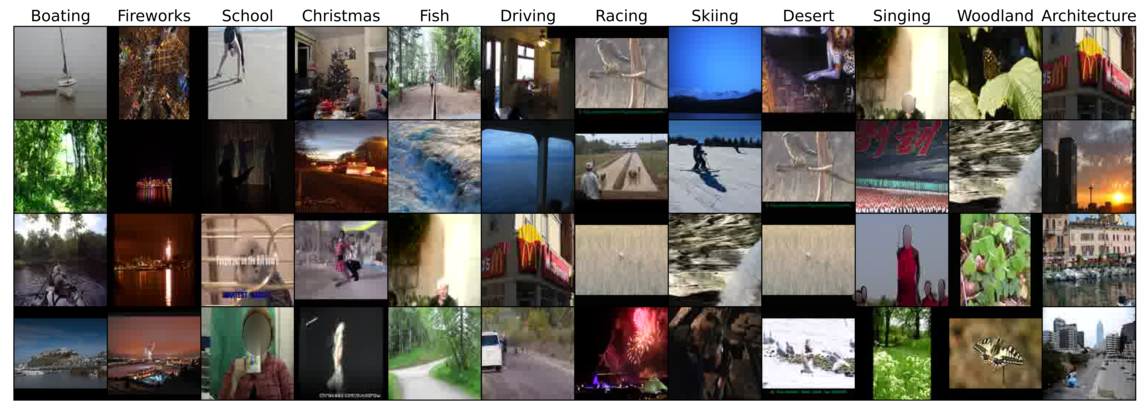}
    \end{center}
  \caption{Frames of videos from various classes.}
  \label{fig:manyFrames}
\end{figure*}

Figure \ref{fig:frames} shows a snapshot of the video frames and their corresponding meta data. The intent behind collecting this information was to make this dataset suitable for tasks beyond action recognition, as evident from the detailed information presented by the meta data about the video in Figure \ref{fig:frames}. Figure \ref{fig:manyFrames} shows frames from videos in various classes. It shows both the diversity present within classes and the kinds of noise that can be present in our dataset.

\begin{table}[t]
\begin{center}
\caption{\small Some statistics from NoisyActions2M.}
\label{tab:datasetStats}
\begin{tabular}{|p{4.5cm}|p{1.5cm}|}
  \hline
  \small Total number of Countries & 212 \\
  \small Average \# videos per Country & 980 \\
  \small Average \# labels per Video & 4 \\
  \small Average Duration & 123s \\
  \small \# videos with duration < 60s & 1723439 \\
  \small \# videos between 60-120s & 423512 \\
  \small Average \# Entities per video & 6 \\
  \small Total \# videos with tags & 968586 \\
  \small Average \# tags per video & 12 \\
  \small Total \# videos with comments & 230269 \\
  \small Average \# comments per video & 5 \\
  \hline
\end{tabular}
\end{center}
\end{table}

Table \ref{tab:datasets} shows a detailed comparison of our dataset with similar video benchmarks. Apart from its scale, one other defining feature of our dataset is the rich meta data information and multi-label information. While Youtube-8M \cite{abu2016youtube} is a much larger multi-label dataset, they only provide frame-level features and don't provide meta data. Similarly, our dataset is much larger than another multi-labeled dataset, HVU \cite{hvu2019} along with meta data. Our dataset is closely related to YFCC100M \cite{thomee2016yfcc100m}. We focus more on action classes and video modality, whereas the YFCC100M dataset focuses more on general entities.

\section{Methodology}

To benchmark a wide variety of learning strategies with a limited compute constraint, we create a 25K and a 100K split of our dataset based on the amount of meta data present for each video. We concatenated all meta data and then picked the top 25K/100K videos where all meta data fields were present and had the maximum number of words. We benchmark various strategies on a 25K split and report results on a 100K split, which subsumes the 25K split.

\subsection{Loss Functions during Pretraining}

We first look at various loss functions for pretraining. We can either do single-label training by selecting a random label for each video or multi-label training. We also devise a loss masking strategy (CCE-Mask), which is essentially single label training, but with masked loss during backpropagation for the ignored labels for a given video. We do this to not send negative feedback for the ignored labels. A detailed algorithm of loss masking is shown in Appendix B.1. For the single label setting, we look at the standard Categorical Cross-Entropy(CCE) loss. For the multi-label setting, we look at Binary Cross-Entropy Loss (BCELoss), Focal Loss (Focal) \cite{lin2017focal}, and Distribution Balanced Focal Loss (Balanced-Focal) \cite{wu2020distribution} since our dataset is heavily imbalanced and long tail with respect to the classes. 

\subsection{Different types of Pretraining}

We also experiment with Self Supervised Learning (SSL) as a pretraining strategy. It has been shown previously \cite{Ghosh_2021_CVPR} that SSL can provide robustness to label noise. Since our dataset can have label noise, we look at two settings: we first pretrain using SSL and then check downstream performance, and the other setting being that we first pretrain using SSL, then pretrain on our dataset with labels and finally check downstream performance, with the idea that performing SSL before will provide robustness against label noise in our dataset.

After benchmarking we run the final methods on the 100K dataset and show their downstream performance on UCF-101 \cite{soomro2012ucf101} and HMDB51 \cite{kuehne2011hmdb}. 

\subsection{Robustness to different types of Corruption}

We use the models trained on the 100K split to demonstrate video corruption and synthetic label noise robustness. For synthetic label noise, we flip a percentage of true labels randomly to some other label. For video corruption, we either randomly flip each bit of the video with a given probability (Random Corruption) or select a random contiguous segment of the video and flip its bits (Contiguous Corruption). The details for Synthetic Noise and Video Corruption experiments can be found in Appendix B.2.

\section{Experimental Setup}

The benchmarking experiments with 25K split are done using the 3D-Resnet-18 (R3D) architecture \cite{hara3dcnns}. For the final experiments on the 100K splits, we use the R(2+1)D-18 \cite{tran2018closer} architecture since it performs better than 3D-Resnets, but at the cost of more GPU memory as shown by the results in the original paper \cite{tran2018closer}. For SSL, we use the Pretext Contrastive Learning setup \cite{tao2020pretext}.

We use the SGD optimizer and Nesterov Momentum with an initial learning rate of $0.03$ and a cosine learning rate scheduler for all experiments. To train models faster on a single GPU, we also use mixed-precision training \cite{micikevicius2017mixed}. We also employ random skips to cover a good portion of the video during loading videos during pretraining, given our constraint of only using $16$ frames from each video. For fine-tuning experiments, however, we use a skip rate of 1. During testing, we average out the results over all clips from a given test video. We use a multi-scale random crop of $112$x$112$ on frames with a horizontal flip.

All models are pretrained for 50 epochs and all fine-tuning experiments were done for 100 epochs, and the best validation score is reported. A batch size of $64$ is used for all experiments. We use split-1 of UCF101 and HMDB51 to report all our fine-tuning results. We use top-1 accuracy. Unless stated, the results for any experiment setup have been reported on split-1 of UCF101.

\section{Results and Discussion}

\subsection{Loss Functions}

\begin{table}[t]
\begin{center}
\caption{\small Results of using various loss functions during pretraining with 25K set and finetuning on UCF101.}
\label{tab:resultsLossFunctions}
\begin{tabular}{|p{2.5cm}|p{2cm}|p{1.5cm}|}
  \hline
  \small Pretraining & Loss Function & Accuracy
  \\
  \hline
  \small None & - & 44.54 \\
  \small Single Label & CCE & 39.78 \\
  \small Single Label & CCE-Mask & 40.73 \\
  \small Multi Label & BCELoss & 44.77 \\
  \small Multi Label & Focal & 45.04 \\
  \small Multi Label & Balanced-Focal & \textbf{45.38} \\
  \hline
\end{tabular}
\end{center}
\end{table}

We first report the results of using various loss functions during pretraining on our 25K split. All results are obtained by first pretraining on 3D-Resnet-18 and then finetuning on UCF101. In Table \ref{tab:resultsLossFunctions} we see that Distribution Balanced Focal Loss (Balanced-Focal) multi-label pretraining performs the best, while single label pretraining performs the worst. Single label training overfits very quickly, which might be why it has the least accuracy. To test this observation, we also finetuned single label pretrained models at different epoch steps (5, 50). We observed that the epoch 5 model got $44.54$ and the epoch 50 model got $38.94$ accuracy, which is roughly similar to the results when we don't pretrain and, in the other case, worse than scratch training. Loss Masking (CCE-Mask) helps, but it's still worse than scratch accuracy. Further, it seems that better multi-label training methods on our noisy and imbalanced dataset help downstream accuracy. 

\subsection{Pretraining Strategies}

\begin{table}[t]
\begin{center}
\caption{\small Results of using various pretraining strategies and their combinations with 25K split and finetuned on UCF101. MLP-BF: Multi Label Pretraining with Balanced-Focal loss.}
\label{tab:resultsPretraining}
\begin{tabular}{|p{4.5cm}|p{1.5cm}|}
  \hline
  \small Pretraining strategy & Accuracy
  \\
  \hline
  \small None & 44.54 \\
  \small SSL-UCF & 49.17 \\
  \small SSL-Kinetics & \textbf{53.21} \\
  \small SSL-25k & {51.57} \\
  \small SSL-25k + MLP-BF & {51.73} \\
  \hline
\end{tabular}
\end{center}
\end{table}

Next, we report results on various types of pretraining, simple, SSL, and their combinations. We first do SSL on UCF101 itself to set a baseline (SSL-UCF). Further, for better comparison, we create a similar-sized split of Kinetics by randomly sampling $26392$ videos for $400$ classes and perform SSL with this dataset (SSL-Kinetics). Finally, we perform SSL with our 25K split (SSL-25K). We observe that results on the model trained with our split perform better than SSL on UCF, but the model pretrained with Kinetics using SSL performs the best. This is expected since the Kinetics videos are very clean and much more correlated with the action labels to which they are assigned. One may argue that SSL does not use label information, but the presence of clean and informative videos seems to impact downstream performance. Balanced-Focal training on top of SSL does not seem to help the final downstream accuracy. Also, these results demonstrate that our noisy videos help with downstream tasks, and its effectiveness is not far from manually curated datasets like Kinetics.

\begin{table}[t]
\begin{center}
\caption{\small Results on the 100k split using R2P1D.}
\label{tab:resultsR2p1d}
\begin{tabular}{|p{3cm}|p{1.5cm}|p{1.5cm}|}
  \hline
  \small  & \multicolumn{2}{c|}{Accuracy}
  \\
  \cline{2-3}
   Pretraining & UCF101 & HMDB51 \\
   \hline
  \small None & 44.81 & 18.24 \\
  \small SSL-100K & \textbf{61.7} & \textbf{29.15} \\
  \hline
\end{tabular}
\end{center}
\end{table}

Finally, we train R(2+1)D models on the 100K split, choosing the best methods from the benchmarking experiments on 25K. The results are shown in Table \ref{tab:resultsR2p1d}, and we can observe that we get a significant boost in accuracy by using a model which has been trained using SSL on 100k (SSL-100K).

\subsection{Robustness Experiments}

\begin{table}[t]
\begin{center}
\caption{\small Synthetic Label Noise Results on UCF101 for the 100k split using R2P1D}
\label{tab:labelCorruption}
\begin{tabular}{|p{3cm}|p{1.5cm}|p{1.5cm}|}
  \hline
  \small Label Noise Percentage & W/o Pretraining & With Pretraining
  \\
  \hline
  \small Asymmetric 40\% & 29.32 & \textbf{45.49} \\
  \small Asymmetric 80\% & 12.74 & \textbf{21.91} \\
  \hline
\end{tabular}
\end{center}
\end{table}

\begin{table}[t]
\begin{center}
\caption{\small Video Corruption Results on the 100k split using R2P1D.}
\label{tab:corruption}
\begin{tabular}{|p{2.8cm}|p{2.3cm}|p{2.3cm}|}
  \hline
  \small Corruption & W/o Pretraining & With Pretraining
  \\
  \hline
  \small Random Corruption & 35.16 & \textbf{45.28} \\
  \small Contiguous Corruption & 42.69 & \textbf{56.46} \\
  \hline
\end{tabular}
\end{center}
\end{table}

Next, we show the results on UCF101 with synthetic label corruption. Table \ref{tab:labelCorruption} shows those results with $2$ different label corruption percentages, with accuracy on the scratch network and accuracy after fine-tuning an R(2+1)D network pretrained with SSL on the 100k. We see significant improvements for both levels of synthetic noise, and even for an extreme synthetic noise percentage of $80\%$, we see an improvement of more than $8\%$ for R(2+1)D. We attribute this robustness to the variety of examples our network learns from during pretraining. In our experiments with just multi-label pretraining, we observed the same robustness during fine-tuning, but the gap, in that case, was smaller than we get when we use SSL for pretraining. More investigation is needed here, especially on varying noise types (Symmetric, Noise Dependent, etc.).

For video corruption, we see similar trends. While the overall accuracy decreases, using a pretrained model gives us $10\%$ more accuracy for random corruption. For contiguous corruption, the scratch accuracy decreases by $2\%$, but the pretrained network still maintains high accuracy. More investigation is needed to validate whether we get similar robustness behavior with other kinds of video corruption. With this, we can now see that pretraining on a noisy dataset will help with both label and video corruption.


\section{Future Work and Directions}

Having shown our dataset's effectiveness for video action classification, with absolutely no human supervision and verification, we now want to take this opportunity to talk about how our dataset can be used for many other tasks. We also talk about some further work that can be done on top of our current results.

\subsection{Future Work}

We performed an analysis with various loss functions and observed that multi-label learning in noisy datasets works. We hope that this will be a good starting point for future work in noisy multi-label learning. One source of obtaining better or more labels is the user-generated meta tags, title, description, and comments. Efficiently extracting labels from this meta data is a challenge. Our first attempts involved using Wordnet Synsets \cite{miller1995wordnet}. With limited preprocessing, we could extract some new useful labels for a given video, but this came at the cost of getting many redundant and noisy labels, making the whole process noisier than before. Good recovery of these implicit labels using video content and its meta data is an exciting challenge and will augment our above experiments. Given that we achieve very competitive results with just the 100K split, the first experiment will be to scale up this training towards all 2M Ids. 

To collect this dataset, we use labels from all datasets in Table\ref{tab:datasets} as our seed queries. These datasets have primarily been collected from Youtube, and our dataset has only been collected from Flickr. These factors allow another interesting use case for our dataset: studying content diversity and distribution shift across platforms for the same set of class labels.

We do not address the aspect of pretraining in the presence of noisy labels. Learning in the presence of label noise is a fairly well-studied problem in various settings \cite{xiao2015learning, frenay2013classification, patrini2017making, rolnick2017deep, lee2018cleannet, vahdat2017toward}. Weakly supervised settings almost always suffer from the problem of corrupted labels. The structure of noise in web scraped datasets is usually not very clear. Hence, using the standard label-noise tackling methods is challenging in such situations and a very interesting research direction. We hope that our dataset acts as a benchmark for such methods.

\subsection{Future Directions}

This dataset intends to set a standard benchmark to learn from noisy web data in various multimedia tasks. It can be used for pretraining or directly for cross-media retrieval tasks since we have videos, video frames (images), meta data (tags, comments, description), and video audio. Our dataset can act as a rich source of pretraining for a variety of different multimedia problems. Some direct applications are Webly supervised cross-modal retrieval, Multimodal video understanding, Action localization, Video Captioning, Video Understanding, No-audio Multimodal Speech detection, Image Captioning, Image Description, and Multimedia Recommender systems. Various statistics about the dataset computed in Section 3 further support the above speculation on future usage.

Our dataset was collected using class labels from well-known datasets shown in Table \ref{tab:datasets}. Due to this, subsets of our dataset can be used as web scraped alternatives to those datasets, allowing for some interesting experiments involving a comparison between web data (real-time distribution) and carefully annotated and curated data for the same set of classes. 

Finally, inspired by results in Table \ref{tab:corruption} and Table \ref{tab:labelCorruption}, one can also examine how pretraining datasets similar to ours can lead to some downstream label and video corruption robustness. It will also be interesting to investigate how pretraining affects other properties of fine-tuned models like Adversarial Robustness, Domain Shift, Out of Distribution Detection, and Uncertainty Estimation.

\section{Conclusion}

This work proposes a new large-scale benchmark dataset for video understanding from noisy data. The proposed dataset is collected using labels from standard video benchmarks, with useful surrounding meta information and all multi-labels corresponding to each data point without human verification. We demonstrated its usefulness in downstream action recognition tasks on two standard action classification benchmarks, UCF101 and HMDB51, and reported significant gains in top-1 accuracies. We also demonstrated an interesting robustness property against varying asymmetric label noise. We hope that this dataset serves as a benchmark for research in noisy learning for videos and is helpful for various multimedia tasks.

\bibliographystyle{ACM-Reference-Format}
\bibliography{sample-sigconf}


\appendix

\section{Appendix: Dataset Construction}

\subsection{Datasets used for Long Sentence Queries}

ActivityNet Captions \cite{krishna2017dense}, MS COCO \cite{lin2014microsoft}, MSR-VTT \cite{xu2016msr}, Flickr30k Denotations \cite{young2014image}, SBU \cite{ordonez2011im2text}, A2D \cite{xu2015can}, Visual Genome \cite{krishna2017visual}, Conceptual Captions \cite{sharma2018conceptual}, Charades \cite{sigurdsson2016hollywood}, Charades-Ego \cite{sigurdsson2018charades}, OID \cite{levinboim2019quality}, TGIF \cite{li2016tgif}, ActivityNet-Entities \cite{zhou2019grounded}

\subsection{Entity Classification Process}

To better understand the broader distribution of the dataset, we considered the 23 top entities used in the Youtube8M \cite{abu2016youtube} paper to visualize the distribution of their labels. We consider each of the textual metadata attributes and pass it through an NLI-based Zero-Shot multilabel Text Classifier \cite{yin2019benchmarking}. This uses the Bart-large-mnli \cite{lewis2019bart} model with the top 23 entities as labels. We then combine the labels obtained over all the attributes and analyze the distribution. Each video may have multiple higher-level entities.

\section{Appendix: Training Methodology}

\subsection{Loss Masking Algorithm}

Algorithm $1$ shows how loss masking is done during pretraining in our experiment.

\begin{algorithm}
\label{lossMask}
\DontPrintSemicolon

  \SetKwInOut{Input}{input}
  \SetKwInOut{Output}{output}
  \Input{A dataset of videos with multi labels.}
  \Output{A model M}
   \newlist{alphalist}{enumerate}{1}
   \setlist[alphalist,1]{label=\textbf{\alph*.}}
   Randomly initialize parameters of model M.
   
   \While{training epochs are left}
   {
   		\begin{enumerate}
   		    \item Build a single label training dataset from the \newline multi label dataset with partial labels using \newline class thresholds. Iterate through the multi \newline label list for the given video:
   		    \begin{alphalist}
   		    \item If a label has less number of videos \newline assigned than the threshold, simply \newline assign it to the video.
   	        \item Otherwise move to the next label in the multi-label list.
   	        \item If any video gets no label, randomly \newline sample a label from the multi-label list.
   		    \end{alphalist}
   		    \item Assign the other non-selected labels in the \newline multi-label list as partial labels.
   		    \item Update weights of model M using the training \newline dataset by masking cross entropy loss values at \newline partial label indices. 
   		\end{enumerate}
   }
   Return model M

\caption{Training using Loss Masking}
\end{algorithm}

\subsection{Checking for Robustness}

We also show some results on a synthetically noised dataset to demonstrate label noise robustness. For these experiments, given a dataset, we change the labels of $x$\% of examples of a given true label to some other label in the dataset randomly and leave the rest to their original label. We then compare the training results on such a dataset with fine-tuning a 3D-Resnet-18  and an R(2+1)d-D-18 model trained on our dataset. 



Finally, to demonstrate robustness against video corruption, we simulated two types of corruption patterns on the UCF101 videos, random corruption and contiguous corruption \cite{changbeyond}. We varied the proportion of video corruptions by value notated as $p \in [0, 1]$. For random corruptions, we flip each bit independently with probability $p=1e-4$. We replace a random contiguous segment of length $p=0.75$ times the file length with flipped bits for contiguous corruptions. We find that some videos after applying corruption become unplaybale and are not decoded to frames, which is in accordance with the observations reported by Chang et al. \cite{changbeyond}.

\end{document}